\documentclass{article}

\author{\textsc{ Louis Crane}}

 \title{\textsc{  SEARCHING FOR EXTRATERRESTRIAL CIVILIZATIONS
 USING GAMMA RAY TELESCOPES}}

\bigskip

\begin{document}

\maketitle

{\bf ABSTRACT} {\it  We investigate the possibility of searching for black hole starships using very high energy gamma ray telescopes. We find that they would be near the lower threshhold for observability, although there are many unknown factors.}

\bigskip

{\bf I. INTRODUCTION}

\bigskip

It has recently been suggested that the object Oumuamua was in fact a probe sent to our solar system by an extraterrestrial civilization.  The extremely improbable trajectory of this body, with its non-gravitational acceleration [1], makes this suggestion plausible. 

In addition, recent work by the Kepler telescope [2] seems to suggest that as many as one star in five in our galaxy has an earthlike planet circling it.

These facts make it seem more likely that our region in the milky way galaxy is inhabited by advanced alien civilizations some of whom  are actively exploring interstellar space.

In  [3] it was suggested that a starship could be constructed using the Hawking radiation from an artificial nanoscopic black hole. The issue of whether such a ship or such a black hole could eventually be constructed is extremely difficult; it is certainly far out of reach of current technology. The construction of a starship propelled by the radiation of such a black hole involves technical peoblems, such as focussing a gamma ray laser to nuclear dimensions and finding a way to reflect gamma radiation for which current technology has no solution. It is not possible to say whether an advanced civilization could ever solve them, or to describe a black hole starship in any detail.

There ia also the problem that Hawking's computations depend on a semiclassical approximation. A full quantum theory of gravity coupled to matter is not available. We do not know what corrections it might make to our calculations.

The experience of human beings in space to date poses extreme obstacles to space colonization. Only with a much denser power source than anything yet tried would it be possible to shield a habitat from the radiation of space and accelerate it continuously to provide a livable human environment. A black hole can convert matter into energy, so it would be the ultimate power source.  An artificial black hole, although extremely difficult to produce or control, would open possibilities that nothing we can currently conceive would equal. We think that it should be investigated as far as it can be.

Recent advances in observational astronomy, however, have made it possible to search for extraterrestrial black hole starships in our galaxy. Interesting potential candidates have been observed, and tests to see if any of them are actual BHSs are not difficult to propose. In this brief note, we review the characteristics of hypothetical black hole starships which might make them observable, describe the recent discoveries of the high energy observational astronomers which seem like possible candidates for BHSs, and propose practicable tests to see if any of the candidates are actual starships.

\bigskip

{\bf II. HOW WOULD WE SEE A BLACK HOLE STARSHIP?}

\bigskip
The proposal to create starships using small artificial black holes uses the hypothetical phenomenon of Hawking radiation. The smaller the radius of a black hole, the higher its temperature, and the lower its mass. The key observation underlying the proposal is that a ``sweet spot'' radius exists of about 2.8 attometers [3]. Such a BH would have a lifespan of about a century, and emit enough energy to accelerate itself to relativistic velocities in a timespan of a few decade It would have an output of about 15 petawatts. Such a black hole would be hotter than any natural object, and emit particles and gamma rays in the gev- tev scale The Hawking temperature is 16 GeV at the sweet spot, but the temperature of a working starship might be greater. The gamma rays emitted would be of somewhat higher energy. All this depends on the semiclassical picture, so a fully quantum treatment might involve corrections we do not know,  Unfortunately, predicting any of the variables precisely would require detailed knowledge of the design of the black hole generator and starship, which is completely beyond us at our current stage of scientific and industrial development. It is also not possible to discuss the apparatus which would focus the Hawking radiation into a collimated beam, so that the apparent intensity of the beam to a distant observer is difficult to ascertain.

So a BHS would appear as a point source of extremely high energy gamma radiation. We would expect to see it in our own galaxy, because collisions with the 3 degree microwave background would tend to attenuate it, and distant galaxies would represent an earlier time frame in which intelligent life would not yet have appeared.

A starship powered by a black hole or other very high energy source, would exhibit a very special pattern of activity. In the first half of its voyage, it would emit an extremely energetic collimated  beam of radiation away from its destination, which observers in a backward pointing cone would see as a pointlike source of very high energy gamma radiation. This radiation would appear redshifted to such an observer, and the red shift would increase in time.

Midway through its voyage, the ship would turn around and begin deceleration. It would disappear to its previous observers, and a new image would appear in a forward cone. The new observers would see the source as blue shifted, and the shift would decrease over time.

Detecting this distinctive pattern should be feasible.
\bigskip

{\bf III. VERY HIGH ENERGY GAMMA RAY SOURCES WHICH HAVE RECENTLY BEEN DISCOVERED. SOME MYSTERIES}

\bigskip

The astronomical study of very high energy gamma rays was frustratingly slow for many years. Great progress has been made in the last ten years, as a result of increases in computational power and the construction of batteries of multiple Cerenkov telescopes [4] . It has been discovered that there are some hundreds of very high energy gamma ray sources in our galaxy. Some of them seem to be associated with neutron stars or binary stars with infalling matter, but over half of them seem to lack any explanation. 

It is particularly difficult to explain why so many of them have no X ray component.

Another mystery is the overall excess of high energy gamma rays coming from the galactic center. It seems to be far greater than any known source such as pulsars could explain.

It is also believed that the gamma ray sources are connected with the problem of the origin of the very high energy cosmic rays; which are still a puzzle.

The proposal of this paper is to examine the unidentified  pointlike very high energy gamma ray sources in the galaxy as candidate starships.  The concentration of gamma rays near the galactic center is suggestive, because it would be a favorable region for interstellar colonization, given its great density of stars.

\bigskip

{\bf IV DETECTABILITY}

\bigskip

The many variables we do not know about a BHS would all affect its observability from current earthbased telescopes. Nevertheless, a rough calculation suggests that a BHS within 100 lightyears might be observable by the current generation of gamma ray telescopes. Varying our assumptions about the construction of a BHS could extend this to 1000 light years or more.

A BHS at the sweet spot of 2.8 attometers with a beam which subtended 1 square light year in the plane perpendicular to its motion containing the earth would  emit 15.7 petawatts of power. This translates into about $ 6\times 10^{-8} $ ergs per second per square meter. or  $8\times 10^
{-7} $ photons per meter per second, assuming 100 GeV photons, well within limits of current sensitivity.

If it were possible to manage a black hole of smaller radius than the sweet spot, it would be more luminous, and hence visible from farther.

As an aside let us explain to the reader that the current telescopes use the earths atmosphere as primary detector, so they have very large effective cross section so $10^{-7} $ photons per rsquare meter per second is in fact observable.

\bigskip

{\bf V. A NATURAL TEST}

\bigskip

So we have seen that, notwithstanding the many uncertainties in our understanding of the detailed nature of artificial black hole starships, they would have one feature which would distinguish them from any natural phenomenon. They would accelerate up to relativistic velocities in a time frame from years to decades. Again, details of their construction we cannot know do not allow us to pin down these rates more precisely. Two cases would appear. A starship in the boost phase of its trajectory would appear redshifted to us, while a decelerating ship would appear blueshifted. In either case the rate of change of the red or blueshift would be on the order of a percent per year. Furthermore, the redshift would increase with time while the blueshift would decrease.

As we explained above,  an accelerating starship would be visible to an observer behind it, while a decelerating ship would be visible ahead of it.

In addition, operating starships would appear and disappear on a time scale of decades or a century or so. This would also distinguish them from natural sources of radiation.

The pattern of shifts we propose to search for, increasing red, decreasing blue, would be quite distinctive. It is extremely hard to imagine that any natural phenomenon would reproduce it. Natural bodies do not exhibit steady relativistic levels of changes in their velocities or gravitational fields.

Another question about the construction of BHSs we were unable to answer in  [3] is whether it is possible to feed matter in to the black hole in order to keep it at a constant temperatiure. If not, the shift in temperature due to the change in radius of the black hole would be superimposed on the red or blue shift. 

A positive result in a search for the pattern we propose would be one of the great experiments of history. A negative result would still explain enough of the nature of the gamma ray sources to be of scientific value. Aside from the mind boggling quality of the possible result, there is really no reason not to make the measurements. Actually the measurements may already be made so that it would be a question of sorting through data looking for the pattern we propose.

\bigskip

{\bf VI. REFLECTIONS}

\bigskip

The problem of the existence of extremely high energy particles and rays in the universe is an old and deep one. Perhaps the idea that extremely high energy density systems play a role in it is not so implausible. Exotic suggestions such as dark matter radiation have been advanced seriously to explain the apparent excess of high energy gamma radiation, which indicates that it is  a deep problem.

If it is really true that black hole starships, or some similar high energy density drive ships exist, they would play a more important role in the galaxy than would be apparent, because only a small number of their focussed beams would reach us on earth. 

As far as the existence of intelligent life elsewhere in the galaxy is concerned, there is really no reason to rule it out. The answer to Fermi's question might simply have been that they were rather hard to see, paradoxically because their emissions were too energetic.

Many people have made the argument that earth is likely to be rare, because so many coincidences went into making its history. But couldnt we say the same of any human life, or of the natural history of any species? Each coincidence may be rare, but how many possible coincidences are there?

The efforts to detect extraterrestrial civilizations via SETI  are crippled by the fact that ordinary radio transmissions such as those produced on earth are only detectable over a radius of a relatively few light years [5] . The radiation of a starship would be tightly beamed, extremely powerful, and in a very high frequency not common in nature. This would make it more detectable if, as in some recent estimates [6] the nearest civilization were to be at a distance on the order of 1000 ly.

An antimatter starship would have a very different radiation profile from a black hole starship. It would emit gamma rays with double the electron mass and double the proton mass. A preliminary search has been made [7] for an antimatter ship with negative results, but inconclusive ones. As we argued in  [3] a BHS seems more promising, so an analogous search seems reasonable.

To summarize, we are proposing that observational astronomers keep track of the frequency spectrum of relatively low intensity  pointlike galactic sources of very high energy gamma rays over a number of years, looking for  continuous slow changes in their frequency.

Let us think of the consequences of a positive result. The existence of starships would indicate the existence of extraterrestrial life, one of the great goals of astronomy. The detailed characteristics of the acceleration and radiation of a BHS would allow us to reverse engineer a starship to a certain extent. It would also be a strong signal that homo sapiens needs to begin working out the technological problems of artificial black hole production in order to attain its destiny.

Perhaps such a common purpose would give human life a sense of meaning. It might counteract the tendency towards nationalism and xenophobia we are experiencing.

A negative result, in which a thorough scan of the galaxy found no starships, would also have profound implications.  Although a negative outcome would not be certain, it would strongly suggest that starships and interstellar colonization were effectively impossible. Hopefully this would emphasize the importance of preserving our planetary home for the future.

This suggests an alternative answer to Fermi's question: they can't get here, so we should only expect the odd microprobe, such as we may have already seen. A microprobe programmed to check one star after another and send back information in a beam, would be cost effective for a civilization which had concluded that colonization was impracticable.

\bigskip

{\bf ACKNOWLEDGEMENTS} The author wishes to thank Professor Avi Loeb for his encouragement, and David Yetter for technical help and support.

\bigskip
{\bf BIBLIOGRAPHY}

\bigskip

[1] M. Micheli et. al. `` Non-gravitational acceleration in the trajectory of 1I/2017 (Oumuamua)''  Nature 2018

\bigskip

[2] N. M. Batalla, ``Exploring exoplanet populations with NASAs Kepler mission `` PNAS 2014

\bigskip
[3 ] L. Crane and S. westmoreland `` Are Black Hole Starships Possible?''   
 arXiv 0908.1803 2009

see also: Starships and Spinoza arXiv 1001.3887 2010

\bigskip

[4] E. Lorentz and R. Wagner `` Very high energy gamma ray astronomy'' 
 European Physical Journal 2012

\bigskip

[5]  A. Loeb and M. Zaldarriaga ``  Eavesdropping on radio broadcasting from galactic civilizations
 with upcoming observatories for redshifted 21cm radiation''
   J. Cosmology and Astroparticle Physics 2007
  
  \bigskip

[6 ]A. Wandel ``  How far are extraterrestrial intelligence and life after Kepler?
 Acta Astronautica 2017
 
 \bigskip

[7] M. J, Harris ``  Limits from CGRO/EGRET data on the use of antimatter as a power source
 by extraterrestrial civilizations ``, J. Br. Interplanet. Soc. 2001

\end{document}